\documentclass[%
 reprint,
superscriptaddress,
 amsmath,amssymb,
 aps,
floatfix,
]{revtex4-2}

\usepackage{graphicx}
\usepackage{dcolumn}
\usepackage{bm}
\usepackage{hyperref}
\usepackage{cleveref}
\usepackage{siunitx}

\begin{document}

\preprint{APS/123-QED}

\title{Observation of an exceptional point in a non-Hermitian metasurface}

\author{Sang Hyun Park}
\thanks{These authors contributed equally}
\author{Sung-Gyu Lee}
\thanks{These authors contributed equally}
\author{Taewoo Ha}
\author{Sanghyup Lee}
\affiliation{Center for Integrated Nanostructure Physics (CINAP), Institute for Basic Science (IBS), Suwon 16419, Republic of Korea}
\affiliation{Department of Energy Science, Sungkyunkwan University, Suwon 16419, Republic of Korea}%

\author{Soo-Jeong Baek}
\author{Bumki Min}
\affiliation{Department of Mechanical Engineering, Korea Advanced Institute of Science and Technology (KAIST), Daejeon 34141, Republic of Korea}

\author{Shuang Zhang}
\affiliation{School of Physics and Astronomy, University of Birmingham, Birmingham B15 2TT, United Kingdom}%

\author{Mark Lawrence}
\email{markl89@stanford.edu}
\affiliation{Department of Materials Science and Engineering, Stanford University, Stanford, California 94305, United States}

\author{Teun-Teun Kim}
\email{t.kim@skku.edu}
\affiliation{Center for Integrated Nanostructure Physics (CINAP), Institute for Basic Science (IBS), Suwon 16419, Republic of Korea}
\affiliation{Department of Energy Science, Sungkyunkwan University, Suwon 16419, Republic of Korea}%
\date{\today}

\begin{abstract}
Exceptional points, also known as non-Hermitian degeneracies, have been observed in parity-time symmetric metasurfaces as the parity-time symmetry breaking point. However, the parity-time symmetry condition puts constraints on the metasurface parameter space, excluding full examination of the unique properties that stem from an exceptional point. Here, we thus design a general non-Hermitian metasurface with a unit cell containing two orthogonally oriented split-ring-resonators with overlapping resonance but different scattering rates and radiation efficiencies. Such a design grants us full access to the parameter space around the exceptional point. The parameter space around the exceptional point is first examined by varying the incident radiation frequency and coupling between split-ring resonators. We further demonstrate that the exceptional point is also observable by varying the incident radiation frequency along with the incident angle. Through both methods we validate the existence of an exceptional point by observing unique level crossing behavior, eigenstate swapping under encirclement and asymmetric transmission of circularly polarized light.
\end{abstract}

\maketitle


\section{Introduction}

\begin{figure*}
    \centering
    \includegraphics[width=0.8\textwidth]{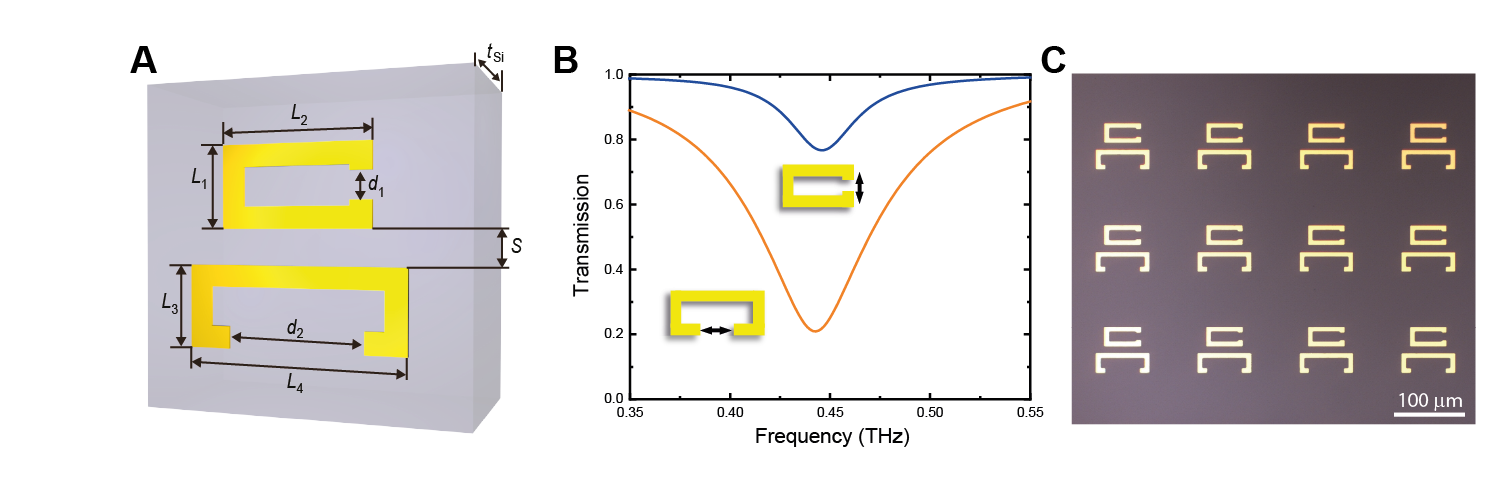}
    \caption{(A) A schematic of the unit cell geometry is shown. The dimensions are set to $L_{1}$=$L_{3}$=30, $L_{2}$=55, $L_{4}$=80, $d_{1}$=10, $d_{2}$=50, $t_{Si}$=525. All units are in $\mu m$. The dimensions of each SRR were optimized such that they have an equal resonance frequency while having different gap sizes. The difference in gap size makes the transmission matrix dispersive, thus letting us control the eigenstates of the system by changing the frequency of incident radiation. (B) Independent transmission plots for the orthogonally oriented SRRs in the unit cell. Each SRR was excited by incident radiation polarized parallel to the gap direction. (C) Top-view microscopic image of the fabricated metasurface.}
    \label{fig:schematic}
\end{figure*}

\par When studying the physics of closed systems in which conservation of energy is strictly obeyed, Hamiltonians are assumed to be Hermitian due to the many appropriate mathematical constraints it puts on the system, such as the realness of eigenvalues, the existence of an orthonormal basis, and unitary time evolution. Many systems in physics are, however, conveniently described by open systems which may either gain or lose energy through interactions with its environment, resulting in complex eigenenergies. Such open systems are represented by non-Hermitian Hamiltonians.
\par One of the most interesting aspects of non-Hermitian(open) systems is their degeneracies, also known as exceptional points(EPs), at which both the complex eigenvalues and eigenvectors of the system coalesce to form a defective eigenspace\cite{Moiseyev2011, Heiss2004, Berry2004, Alu}. Unlike the Dirac point which is unstable under perturbations in two-dimensions, the EP is topologically robust, making it a ubiquitous feature amongst non-Hermitian systems. The significance of exceptional points has frequently been noted in non-Hermitian systems with the special constraint of being parity-time symmetric. In these parity-time symmetric systems the exceptional point was shown to be the point of the parity-time symmetry breaking transition\cite{Feng2017, El-Ganainy2018, Peng2014, Guo2009, Bittner2012, Lawrence2014, Wang2017}. Beyond being the phase transition point of parity-time symmetric systems, exceptional points have unique properties due to the underlying geometry of their complex parameter spaces. Mathematically, the complex parameter space forms a self-intersecting Riemann sheet in which the point of intersection is the exceptional point. The non-trivial topology in the parameter space gives rise to intriguing phenomena such as unconventional level crossing behavior\cite{Heiss1990, Heiss1999}, increased sensitivity to perturbations\cite{Wiersig2016, Chen2017, Hodaei2017} and observation of geometric phase under encirclement\cite{Heiss2012, Dembowski2004, Dembowski2001, Gao2015, Lee2009}. Such behavior has been observed using microwave cavities, chaotic optical cavities, and exciton-polariton billiards.
\par Metasurfaces, two-dimensional arrays of artificially designed subwavelength structures, provide an interesting platform for exploring the physics of EPs. The ability to tune the transmission needed to probe the parameter space in the vicinity of an exceptional point is readily provided in metasurfaces by the variability of array parameters or input frequency. Moreover, measuring the transmission of the metasurface yields \emph{complex} transmission eigenvalues which is crucial to observing the properties of the EP and non-Hermtian matrices. Indeed exceptional points have been observed as a phase transition point in parity-time symmetric metasurfaces \cite{Bittner2012, Lawrence2014}. While the constraints of parity-time symmetry provide a straightforward path for navigating directly through the EP, they also restrict us from exploring the full parameter space around the EP. 
\par In this paper, we demonstrate that topologically robust exceptional points and phase singularities can be observed in a non-Hermitian metasurface. By designing a metasurface composed of hybrid meta-atoms with anisotropic radiation loss, the polarization eigenstates of the metasurface can be manipulated through variation of radiation frequency and meta-atom separation. By analyzing the transmission of the metasurface, we observe phenomena unique to exceptional points including unconventional level crossing behavior, eigenstate swapping under encirclement of the exceptional point and asymmetric transmission of circularly polarized radiation. We have also observed a polarization phase singularity for the first time at an exceptional point in the transmission through an anisotropic metasurface. The increased sensitivity to perturbations near the exceptional point implies that the abrupt phase change observed at the singularity may be used in sensing applications\cite{Kravets:2013csa, :2018hl, Yesilkoy:2018iv}. Finally, we demonstrate that the EP can be observed without repeated fabrication of samples by simply altering the incident angle of radiation. This method provides a path to observing phenomena related to dynamic encircling of the exceptional point\cite{Yoon2018}.

\section{Results and Discussion}
\subsection{Unit cell design}
In this section, we discuss the design of the metasurface unit cell and how it gives a non-Hermitian transmission matrix. The metasurface is comprised of an array with two orthogonally oriented split ring resonators(SRRs) in each unit cell (\cref{fig:schematic}A). The two SRRs in a unit cell are modelled as an effective dipole moment $p_{x,y}=\tilde{p}_{x,y}e^{i\omega t}$ which couple strongly to an incident radiation field $E_{i}=\tilde{E}_{i} e^{i\omega t}$ with a radiative coupling strength $g_{i}$. The geometry and material of the SRR determine the resonance frequency, damping coefficient, and radiative coupling strength of the effective dipole moments. For incident radiation that is close to resonance ($\delta = \omega-\omega_{0} \approx 0 $) and assuming small damping ($\gamma \ll \omega_{0}$) the governing equation for the dipole moments is given by
\begin{equation}
\begin{split}
    \bigg[
    \begin{pmatrix}
    \delta_{x} + i\gamma_{x} & 0 \\
    0 & \delta_{y} + i\gamma_{y}
    \end{pmatrix} 
    &+
    \begin{pmatrix}
    G_{xx} & G_{xy} \\
    G_{yx} & G_{yy}
    \end{pmatrix}
    \bigg] \\ 
    &\times \begin{pmatrix}
    p_{x} \\
    p_{y}
    \end{pmatrix}
    =
    \begin{pmatrix}
    g_{x}E_{ix} \\
    g_{y}E_{iy}
    \end{pmatrix},
\end{split}
\end{equation}
where $G_{ij}$ denotes a summation of the $i$ oriented radiation fields formed by the $j$ oriented dipole moments. Since the coupling terms $G_{ij}$ depend only on the lattice geometry, the use of a square unit cell enforces the conditions $G_{xx}=G_{yy}$ and $G_{xy}=G_{yx}$. Furthermore, since the dipoles are orthogonally oriented and in the same plane, it follows that coupling between the dipoles only occurs via near field quasi-statiradiation, resulting in a real valued $G_{xy}$. While aligning the resonance frequencies of the SRRs is not a necessary condition for observing the exceptional point, in order to maximize the coupling between the two components the resonance frequencies are aligned (\cref{fig:schematic}B). 
\par The transmitted field can then be found by superposing the incident field and the field radiated in the forward direction due to the oscillating dipoles, given by $\frac{i\omega\eta_{0}}{2a^{2}}(g_{x} p_{x},g_{y} p_{y})$ where $\eta_{0}=\sqrt{\mu_{0}/\epsilon_{0}}$ is the free space impedance and $a$ is the lattice constant. The transmission matrix is then

\begin{equation}\label{eq:transmission}
    T = 
    \begin{pmatrix}
    g_{x}^{2}(\delta + i\gamma_{y} +G_{xx}) & -g_{x}g_{y}G_{xy} \\
    -g_{x}g_{y}G_{xy} & g_{y}^{2}(\delta + i\gamma_{x} +G_{xx})
    \end{pmatrix},
\end{equation}
up to a term proportional to the identity matrix that will have no effect on the eigenstates of the system. Since the transmission matrix is non-Hermitian, we expect to observe non-Hermitian phenomena in the transmission eigenvalues and eigenstates. The eigenstates are Jones vectors which represent polarization states and the corresponding eigenvalues will give the magnitude and phase with which the eigenstates are transmitted. Note that $g_{x}=g_{y}$ makes the matrix dispersionless and parity-time symmetric. Thus it is important to ensure that $g_{x}\neq g_{y}$ which is achieved by making the gaps of the two orthogonally oriented SRRs different. 

\subsection{Methods}
\par The system is first studied numerically using the commercial FEM (finite element method) solver CST microwave studio. We check to find a geometry that gives resonant frequencies in the terahertz range. The unit cell geometry of the metasurface used in the simulations and in subsequent experiments is shown in \cref{fig:schematic}A with the resonant frequency of each SRR shown in \cref{fig:schematic}B. 
\par The fabrication procedure of metasurfaces was achieved by conventional photolithography. A positive photoresist (AZ-GXR-601 (14cp), AZ Electronic Materials) was spincoated onto silicon substrate with resistivity of $\sim$ 10 $\Omega$ cm. Au/Cr (100nm/5nm) was then evaporated and patterned as orthogonally oriented SRRs array structures via a lift-off process. 
\par Transmission spectra of the metasurface are characterized by THz time-domain spectroscopy(THz-TDS). THz-TDS yields both the magnitude and phase of the transmission spectrum in a single measurment giving us easy access to the complex transmission eigenvalues of the metasurface. The THz signal is emitted by a commercial photoconductive antenna (iPCA, BATOP). An electro-optic sampling method with a 1-mm thick ZnTe crystal was used to detect the transmitted THz signals. After taking the Fourier transform of the time signal and normalizing against transmission through the bare silicon substrate, this measurement results in a complex transmission spectrum characterizing the sample. The full 2 by 2 transmission matrix is measured in the linear polarization basis. The polarization of the incident and transmitted fields are fixed using two wire-grid polarizers on each side of the metasurface. Transmission for circularly polarized states is related to transmission of linearly polarized states by
\begin{widetext}
\begin{equation}
    T_{cir} = 
    \begin{pmatrix}
    T_{LL} & T_{LR} \\
    T_{RL} & T_{RR}
    \end{pmatrix}
    = \frac{1}{2}
    \begin{pmatrix}
    (T_{xx} + T_{yy}) + i(T_{xy}-T_{yx}) & (T_{xx} - T_{yy}) - i(T_{xy}+T_{yx}) \\
    (T_{xx} - T_{yy}) + i(T_{xy}+T_{yx}) & (T_{xx} + T_{yy}) - i(T_{xy}-T_{yx})
    \end{pmatrix}
\end{equation}
\end{widetext}
where $T_{ij}$ indicates a transmitted wave with polarization $i$ and incident wave with polarization $j$.

\subsection{Results}
\subsubsection{Eigenvalue surface topology}
\par
\begin{figure}
    \centering
    \includegraphics[width=\linewidth]{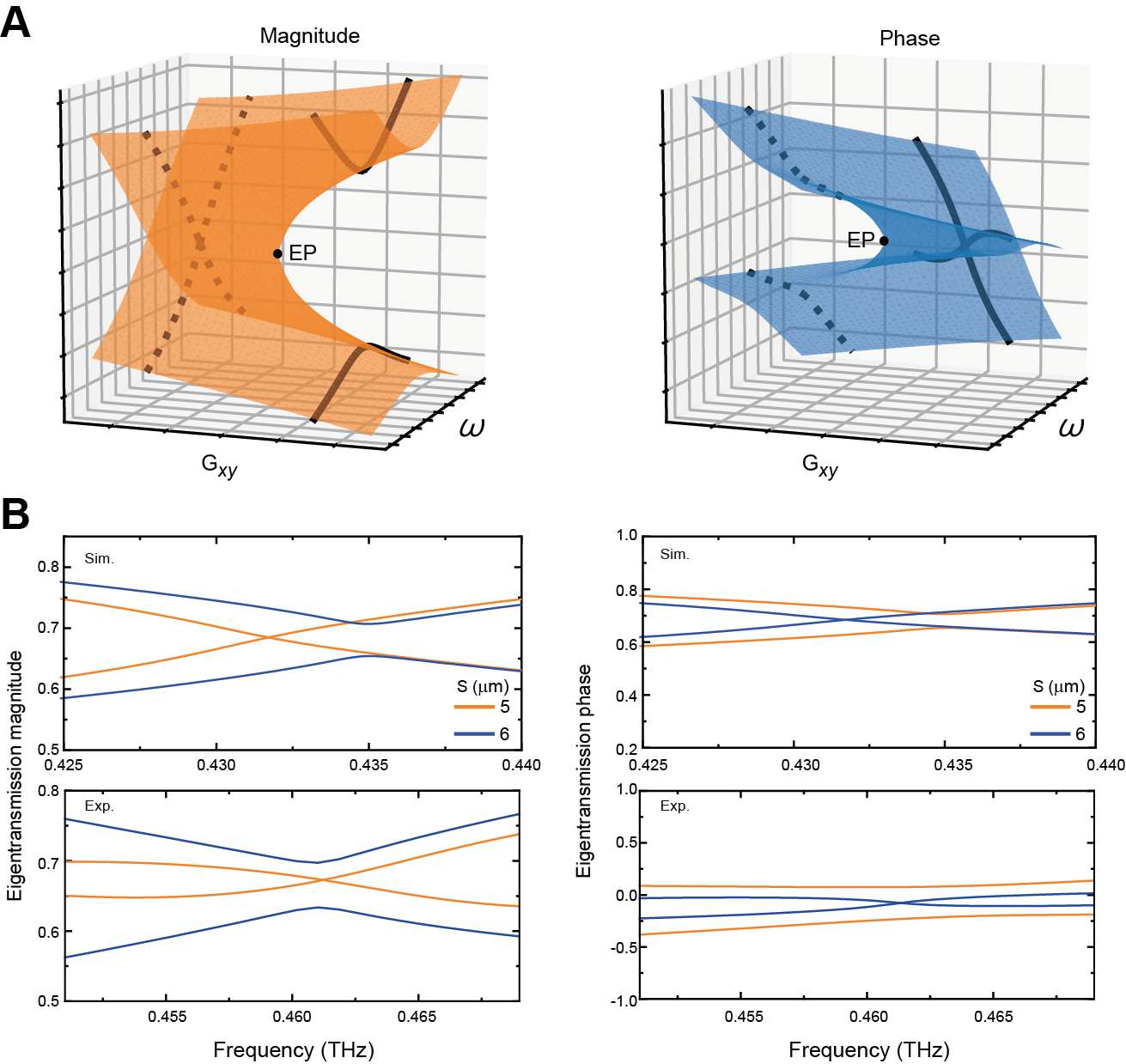}
    \caption{(A) Schematic of the complex eigenvalue surface in $(\omega, G_{xy}(S))$ parameter space. At a given value of $G_{xy}$ we see that the crossing behavior as a function of frequency for the magnitude and phase of the eigenvalues is opposite. As the coupling is varied through the exceptional point the crossing behavior of the magnitude and phase are interchanged. Such behavior is a result of the unique topology of the self-intersecting Riemann surface formed in the vicinity of an exceptional point. (B) Simulation and experimental results of the complex eigenvalues as a function of frequency conducted at different separation, $S$, are shown. When $S$=5 $\mu m$, we see that the magnitude the eigenvalues do not intersect while the phase crosses. When $S$ is changed to 6 $\mu m$, the crossing behavior is swapped and we see crossing in the magnitude and anti-crossing in the phase. As graphically shown in (A), this indicates that an exceptional point exists between $S$=5 $\mu m$ and $S$=6 $\mu m$.}
    \label{fig:riemann}
\end{figure}

 Here we validate the existence of an EP by examining the transmission eigenvalue surface. As mentioned before, the eigenvalue surface of a non-Hermitian matrix forms a self-intersecting Riemann sheet in which the EP is located at the point of intersection(see \cref{fig:riemann}A). This type of topology is unique to non-Hermitian systems and is not observed in Hermitian systems. For Hermitian systems the eigenvalue surface form a double-cone structure with the degeneracy located at the conical intersection. Hence observing the self-intersecting Riemann surface structure will provide strong evidence for observation of an EP.
\par A non-Hermitian degeneracy has a codimension of 2, meaning that we need two parameters to capture the degeneracy\cite{Heiss2004}. Based on the ease and independence with which they can be controlled, the parameters $\omega$ and $G_{xy}$ are chosen. $\omega$ is directly controlled through incident radiation and $G_{xy}$ is controlled by changing the distance between SRRs in the unit cell\cite{Bittner2012, Lawrence2014}. Computing the eigenvalues and eigenvectors of the the transmission matrix given in \cref{eq:transmission}, we find that an exceptional points exist at $(\omega^{EP},G_{xy}^{EP})=\left(\omega_{0},\pm\frac{g_{x}^{2} \gamma_{y}-g_{y}^{2} \gamma_{x}}{2g_{x} g_{y} }\right)$. The eigenstate at each exceptional point is $(1,\pm i)/\sqrt{2}$, which corresponds to left and right circularly polarized light. In this paper, we focus on the exceptional point with left circularly polarized light as its singular eigenstate. 
\par The topology of the self-intersecting Riemann surface can be captured by observing the level crossing behavior in the magnitude and phase of the eigenvalues\cite{Heiss1990, Heiss1999}. While it is usual to examine the real and imaginary parts of the eigenvalues, these quantities lack a direct physical interpretation for the complex transmission eigenvalues. So instead we examine the magnitude and phase of the eigenvalues. Since using the magnitude and phase instead of the real and imaginary parts of the eigenvalues can be considered as a mapping from Cartesian to polar coordinates in the complex plane, the topology near exceptional points will be preserved. When the value of $G_{xy}$ is smaller than $G_{xy}^{EP}$ the magnitude(phase) of the eigenvalues anti-cross(cross). As $G_{xy}$ is increased and passes through the EP, crossing behavior interchanges such that the magnitude(phase) now cross(anti-cross). This type of crossing behavior is unique to non-Hermitian systems and cannot be observed in the double-cone eigenvalue surface of Hermitian systems. 
\par The transmission eigenvalues calculated from simulation and experimental results are shown in \cref{fig:riemann}B. Assuming that the SRR separation distance, $S$, is inversely proportional to the coupling strength ,$G_{xy}$, we observe that the magnitude(phase) of the transmission eigenvalues change from anti-crossing(crossing) to crossing(anti-crossing) as coupling strength is increased. Such behavior exactly follows the topology of the self-intersecting Riemann surface and provides strong evidence that an EP is located in the parameter range $5<S^{EP}<6$.

\subsubsection{Eigenstate swapping}
\par The eigenvalue surface structure shown schematically in \cref{fig:riemann}A also reveals that the EP is a square-root branch-point for the complex eigenvalues. This implies that the eigenvalues of the system will not return to their original values under one full loop around the EP, but will rather be swapped. They will return to their original values only under two full loops. For the eigenstates, the EP is in fact a fourth-root branch point meaning that four full loops are required to return the eigenstates to their initial states\cite{Heiss2004}. A single loop around the EP will interchange the two eigenstates and additionally accumulate a geometric phase of $\pi$ on one of the states. 
\par To show the swapping of eigenstates under encirclement, we follow the evolution of the transmission eigenstates by observing closely spaced stationary eigenstates of the metasurface. The eigenstates of the transmission matrix in \cref{eq:transmission} are polarization states that can be represented as polarization ellipses with ellipticity angle, $\chi$, and orientation angle, $\psi$. Polarization states are plotted on a Poincare sphere for easy visualization. 
\par
\begin{figure}
    \centering
    \includegraphics[width=\linewidth]{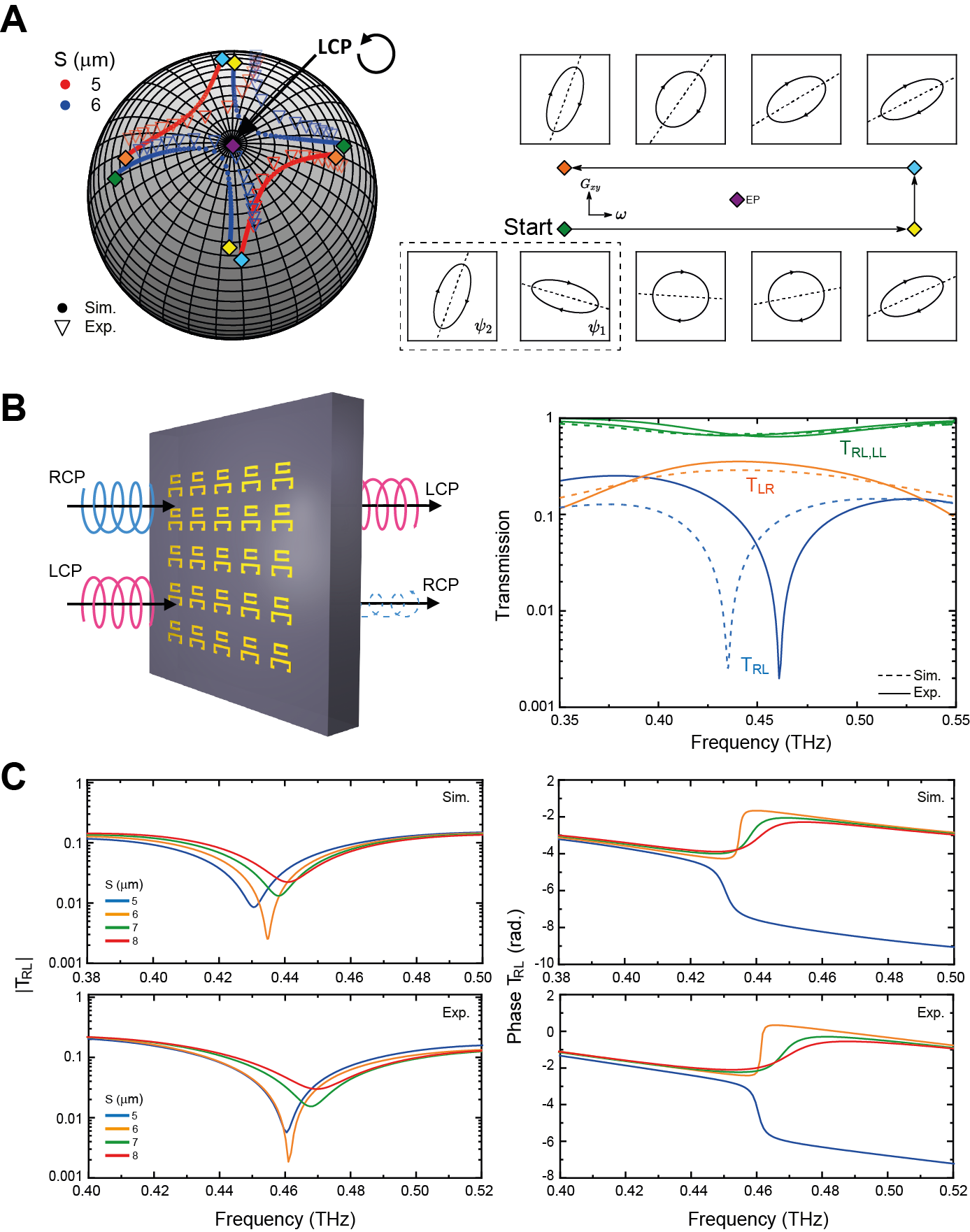}
    \caption{(A) Transmission eigenstates of the metasurface are plotted on a Poincare sphere as the EP is encircled in the $(\omega, G_{xy}(S))$ parameter space. Under one full loop around the EP, the states move half way around the north pole of the Poincare sphere resulting in a swapping of eigenstates (i.e. $(\psi_{1},\ \psi_{2})\rightarrow(\psi_{2}, -\psi_{1})$). To the right of the Poincare sphere we explicitly show the evolution of polarization states. The states in the dashed box show the two eigenstates at the beginning of the loop. Thereafter we follow how state $\psi_{1}$ evolves and find that it becomes $\psi_{2}$ at the end of a single loop. (B) The transmission matrix components in the circularly polarized basis are shown for when S $=6\mu$m. As we have anticipated from \cref{eq:transmission@EP}, $T_{RR}$ and $T_{LL}$ are identical while $T_{RL}$ is suppressed with respect to $T_{LR}$. Asymmetric transmission of circularly polarzied light is a direct consequence of having a single eigenstate at the EP. (C) An abrupt phase flip in $T_{RL}$ is observed as the parameter $S$ is modulated through the EP.}
    \label{fig:sdata}
\end{figure}

 Simulation and experimental data in \cref{fig:sdata}A show how the polarization eigenstates of the metasurface evolve as the parameters $(\omega, G_{xy}(S))$ are encircled around the exceptional point. On the Poincare sphere we plot the eigenstates of the system found from simulations for SRR separations 5 $\mathrm{\mu m}$ and 6 $\mathrm{\mu m} $ and frequency range 0.42 THz to 0.44 THz. Each parameter gives two eigenstates located on opposite sides of the sphere with respect to the north pole. Following the states along the parameter loop $(\omega, S)=(0.42,\ 5) \rightarrow (0.44,\ 5) \rightarrow (0.44,\ 6) \rightarrow (0.42,\ 6) \rightarrow (0.42,\ 5)$, we find that each state moves half way around the north pole resulting in a swapping of eigenstates. The polarization ellipses for the eigenstates along the encircling path are explicitly shown. States from experimental data are also plotted for SRR separations 5 $\mathrm{\mu m}$ and 6 $\mathrm{\mu m} $ and for a frequency range 0.45 THz to 0.47 THz. We find that the states from experimental data also result in a swapping of eigenstates under encircling of the EP. While we should be able to observe an accumulation of geometric phases on one of the states, for our current setup it cannot be measured. This is because measurements for different coupling values are conducted separately, thus making the phase relation between measurements ill-defined.

\subsubsection{Asymmetric transmission of circularly polarized light}

\begin{figure}
    \centering
    \includegraphics[width=\linewidth]{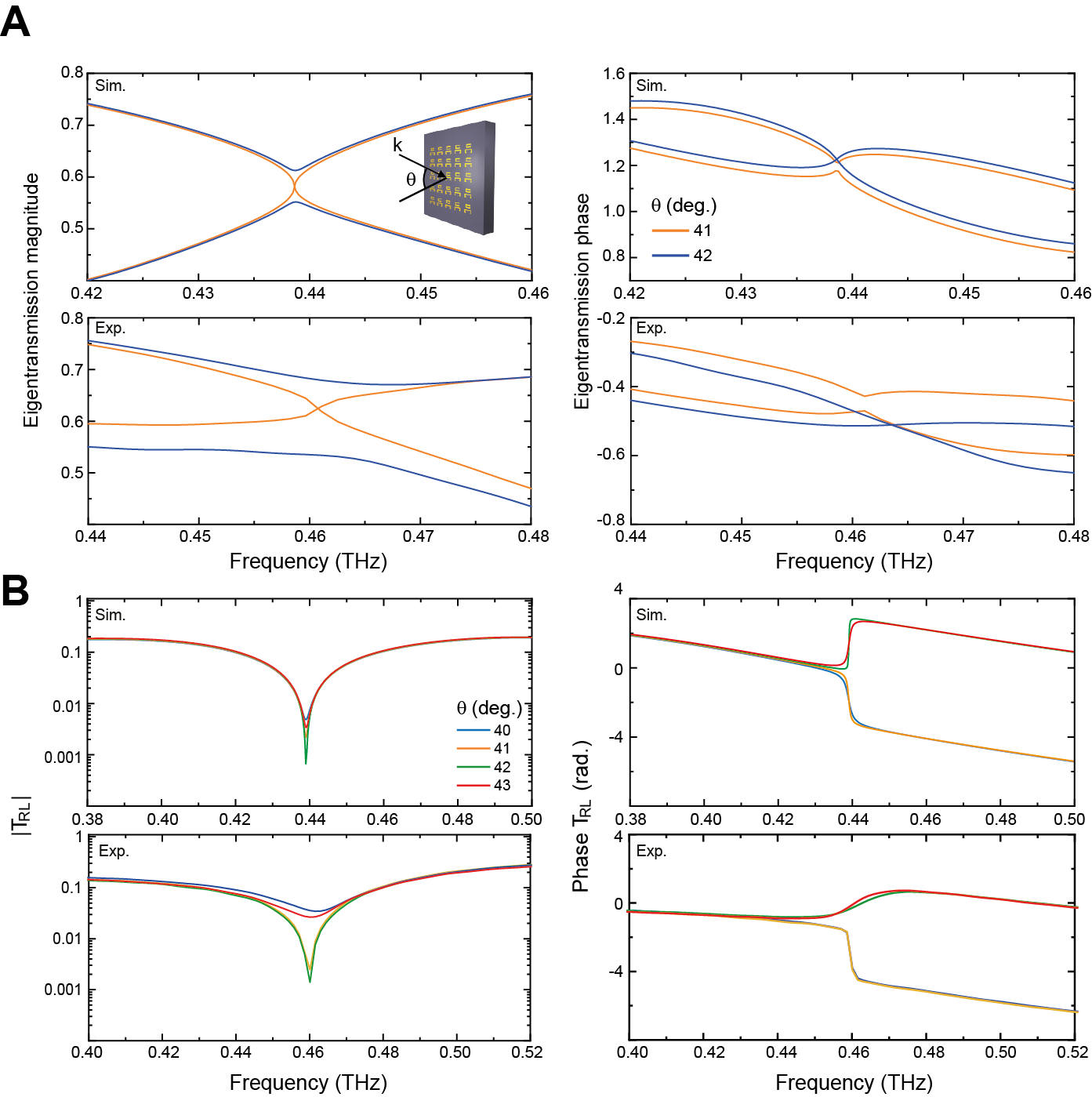}
    \caption{(A) The topology of the self-intersecting Riemann surface is observed in the $(\omega, g_{y}(\theta))$ parameter space. Just as we have observed by varying $G_{xy}$, we see that the crossing behavior of the eigenvalue magnitude and phase are interchanged as $g_{y}$ passes through the exceptional point. From simulation and experimental results we find that the EP located between $\theta=\ang{41}$ and $\ang{42}$. (B) We observe an abrupt phase flip in $T_{RL}$ as $\theta$ is varied through the EP.}
    \label{fig:angledata}
\end{figure}
\par Due to the existence of a single transmission eigenstate at the exceptional point, we expect that transmission should be asymmetric for right-handed circularly polarized (RCP) and left-handed circularly polarized (LCP) states when near the exceptional point, associated with an abrupt change in phase. In general the eigenstates of the system are given by $|\psi_{1} \rangle=\alpha|L\rangle+\beta|R\rangle$ and $|\psi_{2} \rangle=\alpha|L\rangle-\beta|R\rangle$. When the system is near the exceptional point, both eigenstates will be close to the $|L\rangle$ state. Therefore we may assume that $|\alpha|\gg|\beta|$ . Since $T|\psi_{i} \rangle=\lambda_{i} |\psi_{i}\rangle$, we find the transmission behavior for the right and left circularly polarized states to be,
\begin{subequations}\label{eq:transmission@EP}
    \begin{align}
        T|L\rangle = \frac{\lambda_{1}+\lambda_{2}}{2}|L\rangle + \frac{\beta}{2\alpha}(\lambda_{1}-\lambda_{2})|R\rangle \\
        T|R\rangle = \frac{\alpha}{2\beta}(\lambda_{1}-\lambda_{2})|L\rangle + \frac{\lambda_{1}+\lambda_{2}}{2}|R\rangle .
    \end{align}
\end{subequations}
From \eqref{eq:transmission@EP} we can expect that direct transmission of RCP and LCP are identical. However, for cross-polarization conversion, conversion from RCP to LCP is larger than conversion from LCP to RCP by a factor of $(\alpha/\beta)^{2}$. Since we have $|\alpha|\gg|\beta|$ near the EP, we expect to see an asymmetry in the transmission components $T_{RL}$ and $T_{LR}$ as the EP is approached. 
\par We verify these predictions through simulation and experimental results for the transmission spectrum of the metasurface. \cref{fig:sdata}B shows that conversion from LCP to RCP ($T_{RL}$) is significantly lower than conversion from RCP to LCP ($T_{LR}$) when the system is near the exceptional point ($S$ = 6 $\mu m$). Physically, this can be understood from the fact that near the exceptional point, the amplitude of the RCP state is much smaller than the amplitude of the LCP state. Hence, it is relatively improbable for conversion from a high amplitude state to a low amplitude state to occur compared to conversion in the opposite direction. Since the LCP state dominates more as the system approaches the exceptional point, we expect the transmission asymmetry to also get larger. We notice that there is a mismatch of about 0.03 THz in the resonance peak between the simulation and experimental data. This discrepancy most likely stems from a difference in refractive index of the silicon wafer substrate. While we do not dismiss the discrepancy as trivial, we would like to point out that it is not a major issue in observing an EP. The shift in resonance frequency will simply result in the EP also being shifted in the parameter space. One can see that as the system approaches the EP, $T_{RL}$ reached almost zero and an abrupt jump in the phase of $T_{RL}$ is observed despite a small change of coupling strength between two SRRs (\cref{fig:sdata}C). This rapid phase dispersion can be used for extremely sensitive biochemical detection\cite{Kravets:2013csa, :2018hl, Yesilkoy:2018iv}.

\subsubsection{Modulation of incident angle of radiation}
\par Up to this point we have shown that an exceptional point can be observed in our metasurface design through variation of the incident radiation frequency and separation between SRRs. While this provides a straightforward and intuitive path to observing the exceptional point, having to repeatedly fabricate new samples to change the SRR separation is a definite drawback. Repeated fabrication will inadvertently add fabrication errors making it difficult to fine tune parameters. When parameters are near the EP, errors are further amplified because the eigenvalues exhibit an enhanced sensitivity to perturbations\cite{Chen2017, Wiersig2016}. Moreover, we are also unable to perform dynamic encircling of the exceptional point which give rise to phenomena such as time-asymmmetric loops around the EP\cite{Yoon2018}.
\par To overcome these limitations, we modulate the parameter $g_{y}$ instead of $G_{xy}$ by controlling the angle of incident radiation. By increasing the incident angle, we leave the in-plane $x$-component of the field unchanged while weakening the in-plane $y$-component and hence decreaseing $g_{y}$. Using this method, we are capable of probing the parameter space with a single metasurface sample, freeing us from the limitations pointed out above.
\par We now show that the EP is observable in the $(\omega, g_{y}(\theta))$ parameter space. In \cref{fig:angledata}A, simulation and experimental data for the complex eigenvalues are shown. Just as we have seen by modulating the coupling strength, the magnitude(phase) of the eigenvalues change from crossing(anti-crossing) to anti-crossing(crossing) as the incident angle is modulated through the exceptional point. This implies that the self-intersecting Riemann surface topology is also captured in the $(\omega, g_{y}(\theta))$ parameter space. \cref{fig:angledata}B shows the magnitude and phase of the $T_{RL}$ transmission matrix component. Once again, as expected from the properties of an EP, we see that the magnitude of $T_{RL}$ decreases as the exceptional point is approached accompanied with an abrupt phase jump. 

\section{Conclusion}
\par In this work we have designed a non-Hermitian metasurface comprised of orthogonally oriented SRRs to observe an exceptional point. By first controlling the radiation frequency and coupling between SRR elements we have observed phenomena stemming from an EP such as unique level crossing behavior, swapping of eigenstates under encirclement of the EP, asymmetric transmission of circularly polarized light and an abrupt phase flip in the cross-polarization transmission. This abrupt phase change for circularly polarized light could have applications in devices including active polarization modulator and sensitive biosensors. Our work also highlights that the EP can be observed without repeated fabrication of the metasurface by simply varying the incident angle of radiation, making it possible to observe phenomena related to dynamically encircling the EP. It would help to precisely locate the optimized operation point even after the sample fabrication. By showing that an EP is observable in the polarization space of light we open up a new platform for not only studying non-Hermitian physics but also developing very sensitive THz systems with nontrivial phase response.
\\
\begin{acknowledgements}
This work was supported by the Institute for Basic Science of Korea (IBS-R011-D1). This work was also supported by the National Research Foundation of Korea (NRF) through the government of Korea (MSIP) (Grant Nos. NRF-2017R1A2B3012364, 2017M3C1A3013923, and 2015R1A5A6001948), and the Center for Advanced Meta-Materials (CAMM) funded by Korea Government (MSIP) as Global Frontier Project (NRF-2014M3A6B3063709).
\end{acknowledgements}

\bibliographystyle{dgruyter}
\bibliography{writeup_arxiv}

\end{document}